\documentclass[aps,prb,twocolumn,showpacs,superscriptaddress,groupedaddress,floatfix,longbibliography]{revtex4-2}
\usepackage{color}
\usepackage{mathtools}
\usepackage{latexsym}
\usepackage{graphicx}
\usepackage{float}
\usepackage{dcolumn}
\usepackage{bm}
\usepackage{amssymb}
\usepackage{amsmath}
\usepackage{mathtools}
\usepackage[space]{grffile}
\usepackage{xcolor}

\usepackage[colorinlistoftodos]{todonotes}

\begin{document}

\title{Orbital magnetization of Floquet topological systems}
\author{Gabriel E.~Topp$^{1}$} \author{P\"aivi T\"orm\"a$^{1}$}
\author{Dante M.~Kennes$^{2,3}$}
\author{Aditi Mitra$^{4}$} \affiliation{$^{1}$ Department of Applied Physics, Aalto University, FI-00076 Aalto, Finland\\$^{2}$ Institute for Theory of Statistical Physics, RWTH Aachen University, and JARA Fundamentals of Future Information Technology, 52062 Aachen, Germany\\$^{3}$Max Planck Institute for the Structure and Dynamics of Matter, Center for Free Electron Laser Science, Luruper Chaussee 149, 22761 Hamburg, Germany\\$^{4}$
  Center for Quantum
  Phenomena, Department of Physics, New York University, 726 Broadway,
  New York, NY, 10003, USA}

\begin{abstract}
A general expression for the orbital magnetization of a Floquet system is derived. The expression holds for a clean system, and is valid for any driving protocol, and arbitrary occupation of the bands. The orbital magnetization is shown to be large not only for Chern insulators, but also for anomalous phases where the Chern number does not fully account for the topology. In addition, the orbital magnetization is shown to take significant values both for a thermal equilibrium occupation of the Floquet bands, and for occupations determined by a quantum quench from an initial state with zero orbital magnetization. For the latter case, the orbital magnetization is shown to be highly sensitive to van Hove singularities of the Floquet bands. 
\end{abstract}

\maketitle

\section{Introduction} 
A frontier topic in condensed matter physics and atomic, molecular, and optical systems
is the engineering of new states of matter by
Floquet driving \cite{Eckardt17, OkaRev, PhysRevResearch.1.023031, Sondhi20, RudnerRev, RevModPhys.93.041002}. Such driving can
give rise to myriad topological
\cite{Kitagawa10,Zoller11,Rudner13,Carpentier15,Roy16,Else16b,Kyser-I,Kyser-II,Potter16,Po16,Loss17,Potter17,Morimoto17,Roy17,Roy17a,Fidk19,Cirac20,PhysRevB.100.041103}
and ordered phases \cite{Else16a,Khemani16,Khemani16b,Else17,Yao17,Chandran16,Berdanier18,Natsheh21a,Natsheh21b} that have
no analog in static systems. For example, in two spatial dimensions (2D), the Chern number does not fully characterize the topology
under Floquet driving. An
extreme case is one where the
Chern number is zero,
and yet chiral edge modes exist in the system \cite{Rudner13}. These edge modes can give rise to a
quantized charge pumping \cite{Titum16} and a quantized
orbital magnetization \cite{Nathan17, Nathan21} when the bulk states are fully localized by spatial disorder, with the corresponding
system known as an anomalous Floquet Anderson insulator \cite{Titum16}.

Despite this progress, understanding the linear-response properties for
Floquet topological systems is a largely open question. While many new topological invariants have been constructed for Floquet systems,
how these manifest when external probe fields are applied is mostly unexplored.
As a first step, a topological quantum field theory for the driving protocol of
Ref.~\onlinecite{Rudner13} was derived in Ref.~\onlinecite{Gromov21}, and the appearance of an orbital magnetization as a linear response
to an external magnetic field was shown. 
However the orbital magnetization for general Floquet driving, and when the bulk states are not
fully localized is an open and important question
both from a theoretical point of view,
as well as for experiments where bulk states may be conducting.
In contrast, orbital magnetization of static systems has been actively studied
\cite{Ceresoli05,Ceresoli06,Shi07}, with recent applications to
twisted bilayer materials \cite{kennes_moire_2021,Andrei2021,RODRIGUEZVEGA2021168434}  that have been shown to exhibit large orbital magnetic moments \cite{Machida20, Law20}.

We derive a general formula for the orbital magnetization of 2D Floquet systems in the absence of disorder. Our formula holds for any driving
protocol as well as any filling of the bands. We apply our formula to
graphene under Floquet driving, and we present results for two filling profiles, one where the Floquet states are occupied according
to a thermal distribution, and the second where the occupation of the Floquet states is set by a quantum quench from an initial state
with zero orbital magnetization. We show that the orbital magnetization is large even for anomalous phases where the Chern number $C$ does not capture all the edge modes of the system, including the extreme case of $C=0$. 

\begin{figure*}
    \centering
    \includegraphics[width=\textwidth]{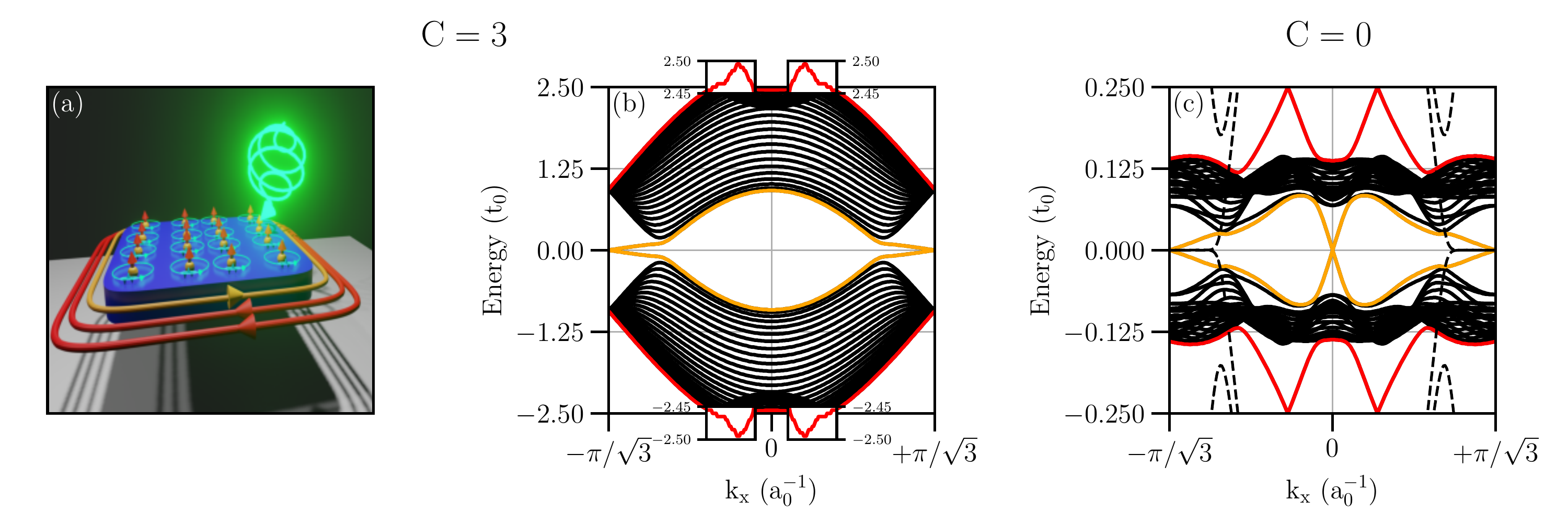}
    \caption{(a) Schematic showing system irradiated by a circularly polarized laser. The yellow (red) loops denote chiral edge modes at Floquet zone center (boundaries). A perturbing perpendicular magnetic field induces an orbital magnetization (tiny loops with arrows indicate induced magnetic moments). (b,c) Floquet bands of driven graphene on a cylinder with periodic boundary conditions in the $x$-direction. The bands have a Chern number of
      (b) $C=3$, and (c) $C=0$. Edge states at the Floquet zone center (boundary) are indicated by yellow (red) color. (b) The insets show two edge states at the Floquet zone boundary (red), and one edge state
      at the Floquet zone center (yellow) (see schematic in (a)). The chirality of the edge modes change from $2$ to $-1$ from
      the zone center to zone boundary consistent with $C=2-(-1)=3$. (c) The zone center and the zone boundary each host two edge
      modes, with no change in the chirality of the edge modes from the zone center to the zone boundary, consistent with $C=0$.
      In (c), for orientation, the bands of static graphene are shown as black dashed lines.
      We take 50 sites (b,c) in the $y$-direction and  keep Floquet harmonics up to
      (b) $|m_{\rm max}|=5$ (c) $|m_{\rm max}|=20$. }
    \label{fig:1}
\end{figure*}

The schematic of the setup is shown in Fig.~\ref{fig:1}(a). A Floquet system with broken time reversal symmetry can host chiral edge modes that traverse the Floquet zone centers  ($\epsilon \approx 0$, yellow) and zone boundaries ($\epsilon \approx \pi/T$, red), with $\epsilon$ being the quasi-energy and $T$ being the period of the drive. We study the response of such a system to an infinitesimal external magnetic field, under spatially periodic boundary conditions, and show that such a probe induces a significant orbital magnetization.

The outline of the paper is as follows. In Section \ref{system} we describe the setup and derive the expression for the orbital magnetization.
In Section \ref{applications} we apply the results to three different topological phases of periodically driven graphene. In addition we present the results for each topological phase for two different occupation probabilities of the Floquet bands.
We present our conclusions in Section \ref{conclusions}, and provide Intermediate steps in the derivation of the formulas, and additional plots in the Appendices. 

\section{System and derivation} \label{system}

Let us consider a general spatially periodic and time periodic Hamiltonian in 2D. If  ${\bf k}$ is
the quasi-momentum, and $T$ is the period of the drive, the Hamiltonian may be written as $H({\bf k}, t)$, where  $H({\bf k},t+T) = H({\bf k},t)$.
According to Floquet theory \cite{Shirley65,Sambe73}, the Floquet eigenstate $|\psi_{n, \bf{k}}(t)\rangle$ can be decomposed
as $|\psi_{n, \bf{k}}(t)\rangle = e^{-i\epsilon_{n, \bf{k}}t}|\phi_{n, \bf{k}}(t)\rangle$ where
$|\phi_{n,\bf{k}}(t)\rangle$ is the time-periodic Floquet quasimode, and $\epsilon_{n,\bf{k}}$ is the quasi-energy, with
$n$ labeling the bands. The first Floquet Brillouin zone corresponds to $ \epsilon_{n,{\bf k}} \in \left[-\pi/T, \pi/T\right]$. 
It is convenient to define the Floquet Hamiltonian
$H_F = H({\bf k},t) - i\partial_t$, which obeys $H_F |\phi_{n,{\bf k}}(t)\rangle = \epsilon_{n,{\bf k}}|\phi_{n,\bf{k}}(t)\rangle$ .

Let us assume that the system has reached a steady-state characterized by
an occupation  $f_{n,{\bf k}}$ of the Floquet bands \cite{Dehghani14,Dehghani15a, Dehghani16b, Esin18}. Thus the system is described by a density matrix 
$\rho(t) = \sum_{n, {\bf k}}f_{n,{\bf k}}|\phi_{n,{\bf k}}(t)\rangle \langle \phi_{n,{\bf k}}(t)|$. Note  that due to the time-periodicity of the Floquet quasimodes, the steady-state still involves oscillations that are periodic in time. Depending on the value of $f_{n,{\bf k}}$, we could be studying pure states or mixed states. The density matrix can also have off-diagonal components. However, when evaluating the average of typical observables, the off-diagonal terms lead to rapid oscillations at incommensurate frequencies. A summation over all the rapidly oscillating terms leads to a decay in time. Since we are interested in the steady-state behavior of the expectation value of observables, we will neglect the off-diagonal components of the density matrix. 

We are interested in studying the linear response of the above system to a weak external magnetic field. It was shown that the orbital magnetization in a Floquet eigenstate time-averaged over one drive cycle equals the rate at which the quasi-energy of that eigenstate changes due to the applied magnetic field (see for example \cite{Nathan17}). With this as the starting point, we will compute the change in the average quasi-energy of each Floquet eigenstate due to the applied magnetic field, weighting the result by the occupation probability of each Floquet eigenstate.

The average quasi-energy for a system described by the density matrix $\rho(t) = \sum_{n, {\bf k}}f_{n,{\bf k}}|\phi_{n,{\bf k}}(t)\rangle \langle \phi_{n,{\bf k}}(t)|$ is
\begin{align}
  E= \sum_{n,{\bf k}} f_{n,{\bf k}}\langle \phi_{n,{\bf k}}(t) |H_F| \phi_{n,{\bf k}}(t)\rangle = \sum_{n,{\bf k}} f_{n,{\bf k}}\epsilon_{n,{\bf k}}.
\end{align}
Our goal is to study the linear response of the above system to a small magnetic field ${\bf B}$ applied in the perpendicular $\hat{z}$ direction
and varying slowly with an in-plane  wave-vector ${\bf q} = q \hat{y}$,
${\bf B} = \hat{z} B \cos(q y)$  \cite{Shi07}. Using a gauge where the vector potential is ${\bf A} = -\hat{x} B \sin(q y)/q $,
the perturbing term corresponding to this magnetic field is (we set $e=1,\hbar=1$) (see Appendix \ref{app1})
\begin{align}
  V(t) = \sum_{{\bf p}} {\bf A}_{{\bf p}}\cdot {\bf J}_{-{\bf p}}(t)={\bf A}_q \cdot {\bf J}_{-q}(t) + c.c.,\label{Vdef}
\end{align}
where the current operator is ${\bf J}_{\bf q}(t)$ = $\sum_{\bf k} c^{\dagger}_{\bf {k+q/2}}{\bf v}({\bf k},t) c_{{\bf k-q/2}}$, ${\bf v}({\bf k},t) $=$ \partial_{\bf k}H({\bf k},t)$,
with ${\bf A}_{q} = -\hat{x}B/(2iq)$.
Denoting $\delta |\psi_{n,{\bf  k}}(t)\rangle $ as the change in the eigenstate to $O(V)$, and assuming that the perturbation is switched on at time $t=0$, we have (see Appendix \ref{app1})
\begin{align}
&  \delta |\psi_{n, {\bf k}}(t)\rangle = -i\sum_{n', {\bf k'}}e^{-i\epsilon_{n, {\bf k}}t}|\phi_{n', {\bf k'}}(t)\rangle \int_0^t dt'\nonumber\\
&\times     \langle \phi_{n', {\bf k'}}(t') |V(t')| \phi_{n, {\bf k}}(t')\rangle
  e^{-i\epsilon_{n, {\bf k}}(t'-t)+i \epsilon_{n', {\bf  k'}}(t'-t)}. \label{delpsi}
\end{align}
Since the quasi-energy does not change to $O(V)$ (or equivalently to $O(B)$, c.f. Appendix \ref{app1}), the leading change to the quasimode is
$ \delta |\phi_{n, {\bf k}}(t)\rangle  = e^{i\epsilon_{n, {\bf k}}t}\delta |\psi_{n, {\bf k}}(t)\rangle$. We also assume that
the perturbation is switched on slowly so that it does not change the occupation of the Floquet states. This is also justified from the fact that changing the occupation requires real inelastic processes that move particles from one eigenstate to another. At $O(B)$, only virtual processes are allowed, and therefore to this order, there is no change in the occupation.

Thus the change in the average quasi-energy due to the external magnetic field is
\begin{align}
\!\!  \delta E (t) &= \sum_{n,{\bf k}}f_{n,{\bf k}}\biggl[\langle \phi_{n,{\bf k}}(t)|H_F|\delta \phi_{n,{\bf k}}(t)\rangle + c.c.\biggr]\!\!=\!M(t)B,
\end{align}
where $M(t)$ is the induced magnetization.
We are interested in the limit of long wavelengths, $q \rightarrow 0$, and long times $t \rightarrow \infty$. Even in the long time limit,
the time-periodicity of the Floquet states gives rise to a residual time-dependence to the magnetization. We will explore the magnetization
averaged over one drive cycle i.e, $\overline{M} ={\lim}_{t\rightarrow \infty} \int_t^{t+T}dt'M(t')/T$, where
$\overline{O}$ indicates average of a quantity over one drive cycle.
We introduce the following notation for the $m$-th Fourier component of various matrix elements
\begin{align}
\!\!  \left[\hat{O}\right]^{m}_{n {\bf k},n' {\bf k+q}}\!\!= \frac{1}{T}\int_0^T dt' e^{-i m \Omega t'} \!\!\langle \phi_{n,{\bf k}}(t')
  |\hat{O}(t')| \phi_{n',{\bf k+q}}(t')\rangle.
\end{align}
We find (see Appendix \ref{app1} for intermediate steps)
\begin{align}
&\overline{M}= \lim\limits_{q\rightarrow 0}\frac{1}{2q}\sum_{n,n',{\bf k},m\in {\rm int}}\left(f_{n, {\bf k}}-f_{n', {\bf k+q}}\right)\biggl[
    \left[H_F\right]^{m}_{n {\bf k},n' {\bf k+q}}\nonumber\\
&\times    \left[v^x({\bf k})\right]^{-m}_{n' {\bf k+q},n {\bf k}}
\frac{i}{\epsilon_{n,\bf{k}}+m \Omega - \epsilon_{n',\bf{k+q}}} + c.c.\biggr].\label{Mfl}
\end{align}
Above, $v^x$
is the velocity operator along the $\hat{x}$ direction.
\begin{figure*}
    \centering
    \includegraphics[width=\textwidth]{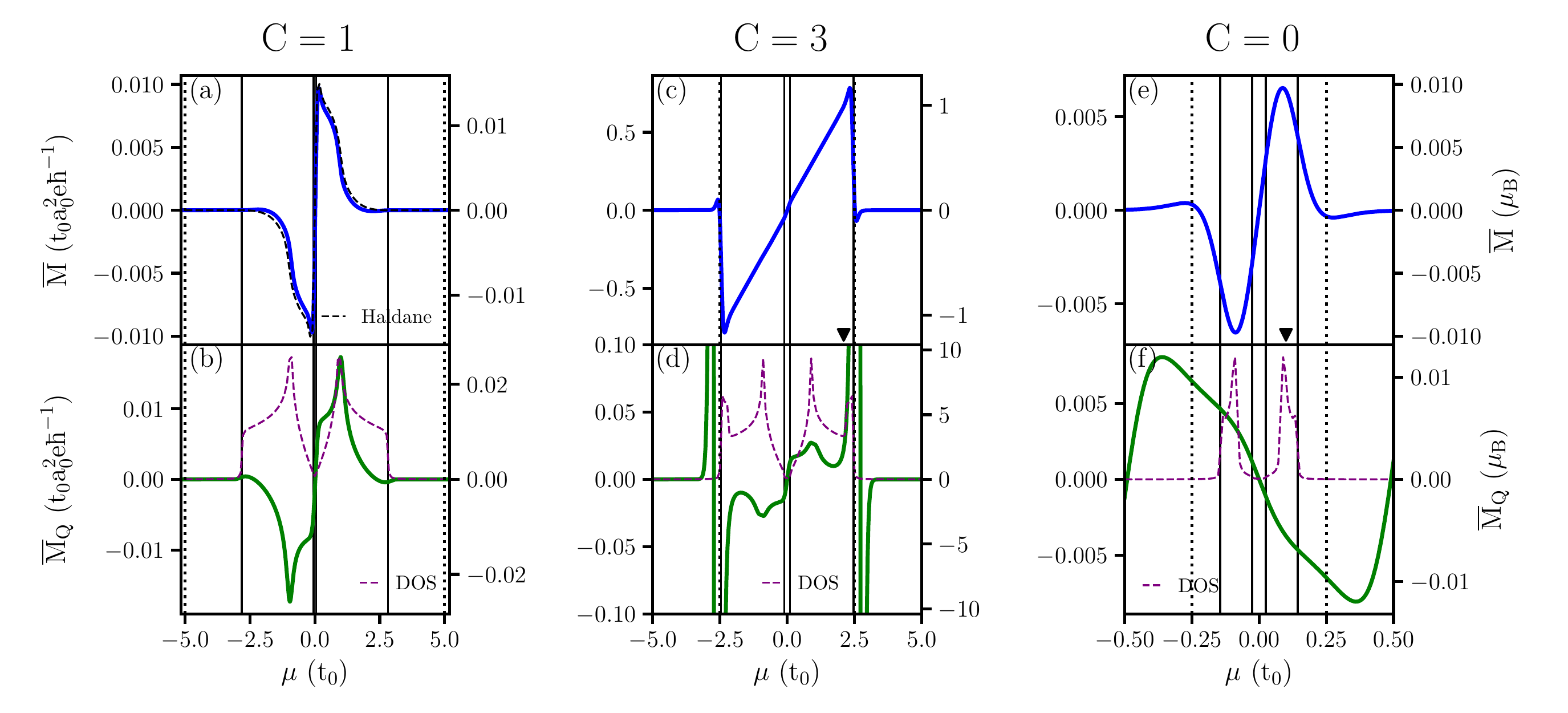}
    \caption{Orbital magnetization per unit dimensionless area $A/a_0^2$ of the spatially periodic system as a function of chemical potential
      $\mu$ for  Chern numbers $C=1,3,0$, with the latter two being anomalous phases.
      Floquet bands filled according to a thermal distribution function (a,c,e) with $\beta^{-1}=0.05t_0$, and according to
      a sudden quench from graphene at
      thermal equilibrium with $\beta^{-1}=0.05t_0$ (b,d,f). 
      Up to $|m_{\rm max}|=3$ (a,b) , $|m_{\rm max}|=5$ (c,d), $|m_{\rm max}|=20$ (e,f) Floquet harmonics were kept. A $1001 \times 1001$ $k$-grid was used for all panels.
      The vertical axes on the right indicate values in units of the Bohr magneton $\mu_B=e\hbar/2m_e$.
      Vertical black dotted lines indicate the Floquet zone boundary $\mu=\pm \Omega/2$. Vertical black solid lines indicate Floquet band edges. a) The black dashed lines indicate reference results 
      obtained from the Haldane model with next-nearest-neighbor hopping $t_2=0.01t_0$, flux $\phi=0.5\pi$, and at the same temperature.
      (c,e) Black triangles indicate the chemical potential values used in Figs.~\ref{fig:3}, \ref{fig:S3}. (b,d,f) the renormalized Floquet DOS is indicated by purple dashed lines (Fig.~\ref{fig:S2} gives more detailes of the DOS). The magnetization changes sign around $\mu=0$ (all panels) and shows strong peaks at the zone boundaries $\mu=\pm \Omega/2$ for the anomalous case $C=3$ (c,d) (see Fig.~\ref{fig:S1} for orbital magnetization over the full range). The orbital magnetization after the quench (b,d,f) is very sensitive to the vHs,  making (d,f) depend on the $k$-grid close to momenta where vHs are found.}
    \label{fig:2}
\end{figure*}

\begin{figure}
    \centering
    \includegraphics[width=\columnwidth]{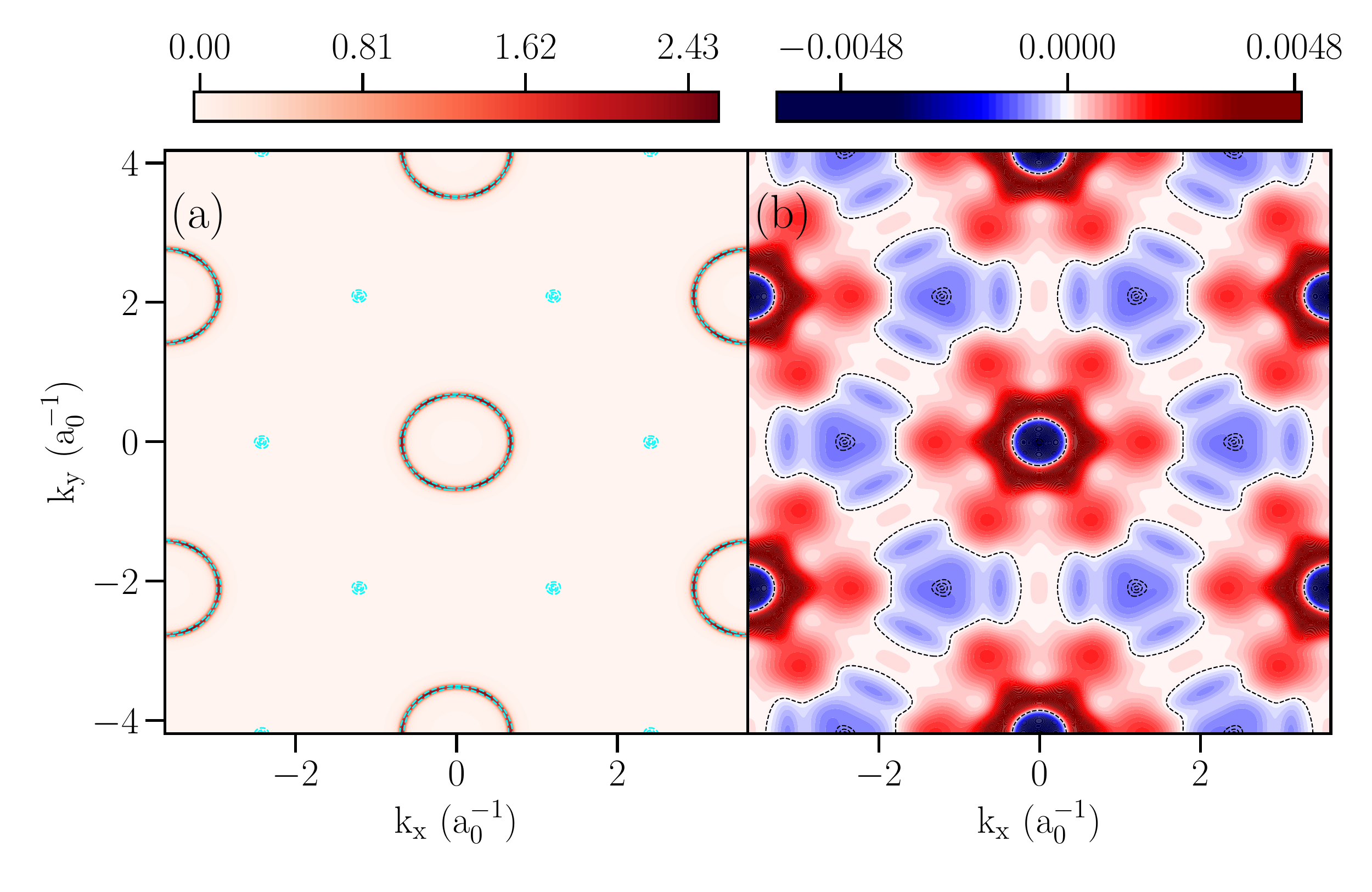}
    \caption{Total magnetization density (integrands of Eq.~\eqref{Meq})
      as a function of quasi-momentum for $C=3$ (a) and $C=0$ (b) and fixed chemical potential $\mu=2.3t_0$
      (a) and $\mu=0.1t_0$ (b) (as indicated by black triangles in Fig.~\ref{fig:2}). Floquet bands are filled according to a thermal
      distribution with
      $\beta^{-1}=0.05t_0$. We employed (a) $|m_{\rm max}|=5$ and (b) $|m_{\rm max}|=20$ Fourier modes and a $1001\times 1001$ $k$-grid.
      (a) Cyan (b) black contours indicate regions where the Berry curvature has significant contributions.}
    \label{fig:3}
\end{figure}

Taking the long wavelength limit is subtle  and the final expression depends on the
distribution function $f_{n,{\bf k}}$. In particular, we find (restoring $e,\hbar$) (see Appendix \ref{app1}),
\begin{align}
  &  \overline{M} =-\frac{e}{2\hbar}{\rm Im}\sum_{n,{\bf k}}\nonumber\\
&\times \biggl[f_{n, {\bf k}}
    \overline{\langle \partial_{{\bf k}} \phi_{n,{\bf k}}(t) |\left(\epsilon_{n, {\bf k}}+H_F\right)\times |\partial_{\bf k}\phi_{n, {\bf k}}(t)\rangle} \nonumber\\
  &-\biggl\{\lim\limits_{q\rightarrow 0}\frac{\left(f_{n, {\bf k}}-f_{n,{\bf k+q}}\right)}{\epsilon_{n,{\bf k}}-\epsilon_{n,{\bf k+q}}}\biggr\}\epsilon_{n,{\bf k}}\nonumber\\
    &\times  \overline{\langle \partial_{\bf k}\phi_{n,{\bf k}}(t) | \left(\epsilon_{n, {\bf k}}- H_F\right)  \times |\partial_{\bf k}\phi_{n, {\bf k}}(t)\rangle}\biggr].
  \label{M}
\end{align}
The cross product indicates that the orbital magnetization depends on the Berry curvature of the bands. 
Since the orbital magnetization is related to the operator ${\bf r} \times {\bf v}$, with ${\bf v} = \partial_t{\bf r} = -i \left[{\bf r}, H(t)\right]$, and since ${\bf r}$ in the momentum basis corresponds to $\partial_{{\bf k}}$, that explains why we have
two partial derivatives in momentum weighted by 
$H_F, \epsilon_{n,{\bf k}}$ \cite{Ceresoli06}.

In order to explore the physics, we need to make some assumptions about the occupations $f_{n,{\bf k}}$. This requires thinking carefully about the initial state before the periodic driving was switched on, and to also account for relevant dissipative processes. When the system is coupled to an ideal reservoir described by a Fermi-Dirac distribution function with a chemical potential $\mu$ and temperature $\beta^{-1}$, and when the Floquet system is driven at frequencies that are large as compared to the bandwidth, it has been shown that the Floquet system acquires the ideal distribution function of the reservoir \cite{Dehghani14, Dehghani15a}. However, for driving frequencies that are resonant, i.e, comparable to the bandwidth, the occupation of the Floquet states can be complicated, and highly dependent on the details of the system-reservoir coupling \cite{Dehghani14,Dehghani15a}. Nevertheless one can imagine performing careful reservoir engineering to obtain desired results for the Floquet occupation.
For this reason, we first consider the case where all states are occupied by a Fermi-Dirac distribution
function at a temperature $\beta^{-1}$, and a chemical potential $\mu$. Moreover, we study corrections to the combination
$E - \mu \langle N\rangle$, where
$N $ is the number operator. This involves shifting $\epsilon_{n,{\bf k}},H_F$ by $\epsilon_{n,{\bf k}} -\mu, H_F-\mu$ thus 
arriving at the following expression for the orbital magnetization 
(see Appendix \ref{app1})
\begin{subequations}
  \label{Meq}
\begin{align}
&\overline{M} = \overline{M}_1 + \overline{M}_2,\\
& \overline{M}_1=-\frac{e}{2\hbar}{\rm Im}\biggl[\sum_{n,{\bf k}}f_{n, {\bf k}}\nonumber\\
&\times    \overline{\langle \partial_{{\bf k}} \phi_{n,{\bf k}}(t) |\left(\epsilon_{n, {\bf k}}+H_F-2\mu\right)\times |\partial_{\bf k}\phi_{n, {\bf k}}(t)\rangle}\biggr], \label{Meq1}\\
  &\overline{M}_2= \frac{e}{2\hbar}{\rm Im}\biggl[\sum_{n,{\bf k}}f'_{n, {\bf k}}\left(\epsilon_{n, {\bf k}}-\mu\right)\nonumber\\
&\times  \overline{\langle \partial_{\bf k}\phi_{n,{\bf k}}(t) | \left(\epsilon_{n, {\bf k}}- H_F\right)  \times |\partial_{\bf k}\phi_{n, {\bf k}}(t)\rangle}\biggr]. \label{Meq2}
\end{align}
\end{subequations}
Above $f'_{n, {\bf k}} = \partial_{\epsilon_{n,{\bf k}}}f$.
Since $\overline{M}_1$ is proportional to the occupation, it survives at zero temperature, while
 $\overline{M}_2$ being proportional to the derivative of the occupation times the energy, contributes only at non-zero temperature.
Eq.~\eqref{Meq} reduces to that for a static system \cite{Shi07} with $H_F$ being replaced by the static Hamiltonian $H$. 
However for static systems in thermal equilibrium,
the orbital magnetization is determined as a linear response
correction to the free energy $F=E-\mu \langle N\rangle - \beta^{-1}S$, $S$ being the entropy and $\beta^{-1}$ being the temperature.
In contrast, here we are considering the linear response of a closed quantum system described by a density matrix $\rho$. For the orbital magnetization, this corresponds to determining corrections to the quasi-energy of each Floquet eigenstate weighted by the occupation of that state \cite{Nathan17}. Thus the natural object that appears here is corrections to $E$ or $E-\mu \langle N\rangle$, where $\mu \langle N\rangle $ is a convenient shift.

For a two-band Floquet system with particle-hole symmetry, the orbital magnetization per unit area $A$ when one of the bands is fully occupied and the other is empty, reduces to   $\overline{M}/A = -(e/\hbar)C \mu/2\pi$ (see Appendix \ref{app1}). Thus the magnetization vanishes for this ideal filling for anomalous phases with $C=0$.
Therefore, in order to probe the $C=0$ phase, we need to either break particle-hole symmetry, or raise the temperature in order to occupy the other band, or simply have a nonequilibrium occupation of the Floquet bands.

As discussed above, the occupation of Floquet states is not guaranteed to be in thermal equilibrium even when they are coupled to an ideal reservoir
\cite{Dehghani14, Dehghani15a,Dehghani16b} except for some limiting cases of high frequency driving or for very careful reservoir engineering. The most natural distribution is one arising from
a quantum quench from an initial state where a drive was absent, and then the drive was switched on following a certain protocol.
For a sudden switch-on of the drive at a time $t=0$, the quench distribution function is
\begin{align}
f_{n,{\bf k}} =\sum_{\alpha} |\langle \phi_{n, {\bf k}}(0)|\psi^{\rm in}_{\alpha, {\bf k}}\rangle|^2 f^{\rm in}_{\alpha, {\bf k}}, \label{fq}
\end{align}
where $|\psi^{\rm in}_{\alpha, {\bf k}}\rangle$ are the eigenstates of the Hamiltonian in the absence of drive, and $f^{\rm in}_{\alpha, {\bf k}}$
is the occupation of these states, which we take to be 
a Fermi-Dirac distribution at a temperature $\beta^{-1}$ and a chemical potential $\mu$.

The orbital magnetization following a quantum quench, and  obtained from corrections to the average quasi-energy $E$ is (see Appendix \ref{app1})
\begin{align}
& \overline{M}_Q = -\frac{e}{2\hbar}{\rm Im}\sum_{n,{\bf k}} \biggl[f_{n, {\bf k}}
    \overline{\langle \partial_{{\bf k}} \phi_{n k}(t) |\left(\epsilon_{nk}+H_F\right)
      \times |\partial_{{\bf k}}\phi_{nk}(t)\rangle}\nonumber\\
    &- \biggl\{ \sum_{\alpha} |\langle \phi_{n, {\bf k}}|\psi^{\rm in}_{\alpha,{\bf  k}}\rangle|^2 \partial_{\epsilon}f^{\rm in}_{\alpha} \frac{v_y^{\rm in}(\alpha, {\bf k})}{v_y(n,{\bf k})}\nonumber\\
   & + f^{\rm in}_{\alpha,{\bf k}}   \frac{\partial_{k_y}\left(|\langle \phi_{n, {\bf k}}|\psi^{\rm in}_{\alpha, {\bf k}}\rangle|^2\right)}{v_y(n,{\bf k})} \biggr\}\nonumber\\
&\times \epsilon_{n, {\bf k}}
    \overline{\langle \partial_{\bf k}\phi_{n k}(t) | \left(\epsilon_{n, {\bf k}}- H_F\right)\times  |\partial_{\bf k}\phi_{n, {\bf k}}(t)\rangle}\biggr].\label{MQ}
\end{align}
There are two main differences between Eq.~\eqref{MQ} and Eq.~\eqref{Meq}. One is that the occupation probabilities entering
in Eq.~\eqref{MQ} are given by Eq.~\eqref{fq} rather than a Fermi-Dirac distribution function. Second, the term coming from
$\lim_{q\rightarrow 0}\left(f_{n, {\bf k}}-f_{n,{\bf k+q}}\right)/(\epsilon_{n,{\bf k}}-\epsilon_{n,{\bf k+q}})$ (c.f. Eq.~\eqref{M}) is quite subtle for the quench
as the energies entering in $f_{n,{\bf k}}$ are those of the pre-quench Hamiltonian before driving was switched on.
Thus a ratio of velocities of the pre and post
quench systems $v_y^{\rm in}(\alpha, {\bf k})/v_y(n,{\bf k})$, appear in the formula for the quench,
with the orbital magnetization becoming sensitive to
van-Hove singularities (vHs) ($v_y(n,{\bf k}_0)=0$) of the Floquet bands. We emphasize that vHs are always important whenever a sum on all momenta is involved. Thus vHs play a role even for a thermal occupation of the bands. However, as Eq.~\eqref{MQ} shows, the role of vHs, especially those coming from $v_y=0$, become more important for a quench 
occupation of the bands than for a thermal equilibrium occupation of the bands.

\section{Application to driven graphene} \label{applications}

We now apply the above formulas for the orbital magnetization to periodically driven (solid-state or artificial) graphene. We
choose a driving protocol where ${\bf k}a_0 \rightarrow {\bf k}a_0 + A_0a_0 \left(\cos(\Omega) \hat{x} -\sin(\Omega t)\hat{y}\right)$ (corresponding to a circularly polarized laser), where
$a_0$ is the spacing between nearest-neighbor sites, $A_0 a_0$ is the dimensionless drive amplitude, and $\Omega$ is the drive frequency.
The chosen driving protocol effectively breaks time-reversal symmetry, with the high frequency limit
\cite{Oka09,Tanaka10, Kitagawa11, Esslinger14}  corresponding to the Haldane model \cite{Haldane88}. 
Denoting the hopping amplitude of graphene by $t_0$, and the velocity at the Dirac points by $v_F$, the gap at the Dirac points
in the high frequency limit is  $\approx 2 (A_0v_F)^2/\Omega$ \cite{OkaRev}, where $v_F = (3/2) t_0 a_0$ \cite{NetoRev}. 

On varying the drive frequency and amplitude, a rich phase diagram is obtained which includes
conventional Chern insulators with edge modes completely characterized by the Chern number, as well as anomalous phases where
the Chern number is insufficient to characterize the topology \cite{Kundu14,Dehghani15a, Dehghani15b, Dehghani16a, Yates16}.
The bandwidth of graphene is $D \approx 6t_0$. In order to highlight the main physics, we will study three qualitatively different cases.
One is that of high frequency driving $\Omega \gg D$, 
the second is that of a weak $A_0a_0\ll 1 $ but resonant drive $\Omega \lesssim D$, while the third is that of a strong $A_0a_0\gg 1$ and resonant drive $\Omega < D$ where the Floquet bands bear little resemblance to that of graphene.
The high frequency case 
shows physics identical to the Haldane model.
The other two cases, for our parameters, correspond to anomalous phases. 

In particular, the case of weak amplitude but resonant drive
(Fig.~\ref{fig:1}, panel (b) $\Omega = 5 t_0, A_0a_0=0.5 $), has Chern number $C=3$. On a cylinder,
the system hosts 3 chiral edge modes, with two of them  of the same chirality and traversing the Floquet zone boundary (red), while the third of the opposite chirality and traversing the zone center (yellow). The Chern number of $3$ measures the change in the chirality from the zone center to the zone boundary,
but cannot determine the number of edge modes at the two gaps.
The second anomalous phase (Fig.~\ref{fig:1}, panel (c), $\Omega = 0.5 t_0, A_0a_0=10$) is an example of a low frequency and high amplitude drive  for which
$C=0$, but the system still hosts chiral edge modes on the cylinder. In particular, the Floquet zone center (yellow) and boundary (red) each host two chiral edge modes, with all edge modes of the same chirality. 

While Fig.~\ref{fig:1} is for a system with boundaries, the rest of the paper presents results for a spatially periodic system. The results for the orbital magnetization for the three cases are plotted in Fig.~\ref{fig:2}.
The first, second and third columns correspond to $C=1,3,0$ respectively, with results presented in units of $e t_0 a_0^2/\hbar$
on the left axes, and in Bohr magnetons $\mu_B=e\hbar/(2 m_e)$ on the right axes, where $m_e$ is mass of the electron, and $t_0,a_0$ are for solid-state graphene.
We choose a temperature of $\beta^{-1}=0.05 t_0$, with the
Floquet zone boundaries $\pm \Omega/2$ indicated by vertical dashed lines, and the Floquet band edges by vertical solid lines. 
The magnetization $\overline{M}$ for a thermal equilibrium occupation is plotted in the first row, while the second row shows the quench orbital
magnetization $\overline{M}_Q$. 

As expected, the high frequency case of $C=1$, for a thermal occupation of the bands, agrees with the Haldane model (a), with the orbital magnetization showing structure as $\mu$ traverses the gap at zero quasi-energy. The steep linear rise in $\overline{M}$ across $\mu=0$, is proportional to $C$. The
quench magnetization $\overline{M}_Q$ (b), in addition to changing rapidly around $\mu=0$, also shows sharp features
due to the dependence of $\overline{M}_Q$ on vHs (see coincidence with sharp structures in the density of states (DOS), purple dashed lines). 

The $C=3$ case shows the same behavior as $C=1$ close to $\mu=0$ (c), however the largest orbital magnetization appear near the Floquet zone boundaries $\mu=\pm \Omega/2$ (in contrast, the $C=1$ case has zero orbital magnetization near $\mu=\pm \Omega/2$). We associate the large peaks visible in $\overline{M}, \overline{M}_{Q}$ (see Fig.~\ref{fig:S1} for the orbital magnetization over the full range) at $\mu \approx \pm \Omega/2$ for $C=3$ as an example of bulk-boundary correspondence as
two additional edge  modes of the {\bf opposite chirality} appear at $\mu=\pm \Omega/2$
relative to $\mu=0$, enhancing the orbital magnetization. In addition, the sharp features in $\overline{M}_Q$ coincide with the vHs (d).

The final case is $C=0$ (e,f) where the bands have a non-zero
Berry-curvature, although it integrates to zero. The non-zero orbital magnetization arises from the non-zero Berry curvature.
Since the chirality does not change from the Floquet zone centers to the zone boundaries,
the strength of the orbital magnetization is of the same magnitude at these points. Note that a temperature of
$\beta^{-1}=0.05 t_0$ is "high" for this case, as a result the orbital magnetization is smoother, and gives non-zero contributions even when $\mu$ is outside the band edges. Despite the Chern number being zero, the orbital magnetization takes comparable values to the $C=1$ case, making it a good probe of anomalous Floquet phases.

The integrand of Eq.~\eqref{Meq}, namely the orbital magnetization density, is shown in Fig.~\ref{fig:3}
(and for Eq.~\eqref{MQ} in Fig.~\ref{fig:S3}) for a chemical potential where the orbital magnetization takes large values (black triangles
in Fig.~\ref{fig:2} (c,e)). The plots highlight the connection between peaks in the orbital magnetization density and
peaks in the Berry curvature. A non-zero Berry curvature is a prerequisite for a non-zero orbital magnetization density. However, the distribution functions, defined by the chemical potential $\mu$ and the temperature $\beta^{-1}$, determine which parts of the band Berry curvature actually contribute. In Fig.~\ref{fig:3}a the Dirac points (small cyan circles) do not contribute to the magnetization density because a chemical potential of $\mu=2.3t_0$ was chosen so that the Dirac points are fully occupied and the Berry curvature contributions from both Dirac bands cancel out (see also Fig.~\ref{fig:S2}). The area around the $\Gamma$-point (big cyan circle), however, is only partly occupied, which together with the peaks in Berry curvature in this region, lead to a strong magnetic response. The same reasoning applies to Fig.~\ref{fig:3}b where we chose a different chemical potential, $\mu=0.1t_0$.

\section{Conclusions} \label{conclusions}
We derived a general expression for the orbital magnetization of bulk Floquet systems and applied it to periodically driven graphene. The orbital magnetization has a lot more information than the Chern number, showing significant response even for anomalous phases with $C=0$. Thus, orbital magnetization is a more sensitive probe of Floquet induced topology than traditional metrics like the Chern number. Ref.~\onlinecite{Nathan17} showed that when the bulk states are localized, the orbital magnetization is quantized and related to a 3D winding number \cite{Rudner13}. However, in our setting where the bulk states are delocalized, the response is non-universal. In addition, while a $C=0$ phase shows a non-zero orbital magnetization, its time-averaged value does not determine in any obvious way how many edge modes the $C=0$ phase has. In contrast when $C\neq 0$, so that edge modes at the center and boundaries of the Floquet Brillouin zone have opposite chirality, this manifests very clearly in the orbital magnetization (c.f. $C=3$ and $C=1$ in Fig.~\ref{fig:2}). 
Despite this lack of complete information, our results will be useful for experiments in Floquet systems where bulk states are likely to be delocalized. 

In this paper we studied the orbital magnetization averaged over one drive cycle. It will also be useful to study the time-dependent orbital magnetization, and extract its Fourier components. For this case we expect to see large response at values of $\mu$ that coincide with
the vHs  and also its Floquet sidebands i.e., for $\mu = \mu_0, \mu=\mu_0 \pm \Omega, \mu = \mu_0 \pm 2 \Omega \ldots$.

\emph{Acknowledgements:} The authors thank Qian Niu for helpful discussions. This work was supported by the US Department of Energy, Office of
Science, Basic Energy Sciences, under Award No.~DE-SC0010821 (AM), 
by the Deutsche Forschungsgemeinschaft (DFG, German Research Foundation) through RTG 1995 and under Germany's Excellence Strategy - Cluster of Excellence Matter and Light for Quantum Computing (ML4Q) EXC 2004/1 - 390534769 (DMK), and by the Academy of Finland under project numbers 303351 and 327293 (GT and PT). Computing resources were provided by Triton cluster at Aalto University (GT). DMK acknowledges support from the Max Planck-New York City Center for Non-Equilibrium Quantum Phenomena.


%

\begin{widetext}

\appendix

\section{Intermediate steps in the derivation of Eq.~\eqref{M}} \label{app1}

The Floquet quasimodes $|\phi_{nk}(t)\rangle$ are time-periodic $|\phi_{nk}(t)\rangle= |\phi_{nk}(t+T)\rangle$, and therefore can be expanded in
Fourier components as follows
\begin{align}
 |\phi_{nk}(t)\rangle =\sum_{m\in {\rm int}}e^{i m \Omega t} |\phi_{nk}^m\rangle.
\end{align}
The Floquet quasimodes obey 
\begin{align}
  H_F|\phi_{nk}(t)\rangle = \epsilon_{nk}|\phi_{nk}(t)\rangle; \, H_F=H(t) -i\partial_t,
\end{align}
and the  Floquet eigenstates are
\begin{align}
  |\psi_{nk}(t)\rangle = e^{-i\epsilon_{n k}t}|\phi_{nk}(t)\rangle.
\end{align}

Let us define the time-evolution operator for the Floquet system as
\begin{align}
  U(t,0) &= {\cal T}e^{-i \int_0^t dt'\left[ H(t')\right]}.
\end {align}
Thus the Floquet unitary translates the Floquet eigenstate forward in time as follows
\begin{align}
  U(t,0)|\psi_{nk}(0)\rangle &= |\psi_{nk}(t)\rangle = e^{-i\epsilon_{nk}t}|\phi_{nk}(t)\rangle.
\end{align}
The perturbing term $V(t)$ is (setting $e=\hbar=1$)
\begin{align}
    V(t) = \sum_{{\bf p}} {\bf A}_{{\bf p}}\cdot {\bf J}_{-{\bf p}}(t).\label{V1s}
\end{align}
Since we choose a perturbation at a specific wavevector $q$, ${\bf A}({\bf r}) = -\hat{x} B \sin(q y)/q $, the sum on ${\bf p}$ only involves two terms, ${\bf p}=q \hat{y}$ and ${\bf p}=-q \hat{y}$. Thus we arrive at
Eq.~\eqref{Vdef}, which we rewrite below for convenience
\begin{align}
  V(t) = {\bf A}_q \cdot {\bf J}_{-q}(t) + c.c.,
\end{align}
with ${\bf A}_{q} = -\hat{x}B/(2iq)$.

In the presence of a perturbation $V(t)$, the Floquet unitary expanded to $O(V)$ is
\begin{align}
  {\cal T}e^{-i \int_0^t dt'\left[ H(t') + V(t')\right]}&\approx U(t,0) -i {\cal T} \biggl[\int_0^t dt' U(t,t') V(t')U(t',0)\biggr].
\end{align}
The change in the wave-function to leading order in $V$ is
\begin{align}
  \delta |\psi_{n k}(t)\rangle &= -i \int_0^t dt'U(t,t') V(t') U(t',0)|\psi_{nk}(0)\rangle  \nonumber\\
  & = -i\sum_{n' k'}|\psi_{n'k'}(t)\rangle\int_0^t dt' \langle \psi_{n' k'}(t) | U(t,t') V(t') U(t',0)| \psi_{nk}(0)\rangle\nonumber\\
  & = -i\sum_{n' k'}|\psi_{n'k'}(t)\rangle\int_0^t dt' \langle \psi_{n' k'}(0) | U(0,t)U(t,t') V(t') U(t',0)| \psi_{nk}(0)\rangle\nonumber\\
  &= -i\sum_{n' k'}|\psi_{n'k'}(t)\rangle\int_0^t dt' \langle \psi_{n' k'}(0) |U(0,t') V(t') U(t',0)| \psi_{nk}(0)\rangle\nonumber\\
  & =  -i\sum_{n' k'}e^{-i\epsilon_{n' k'}t}|\phi_{n'k'}(t)\rangle\int_0^t dt' \langle \phi_{n' k'}(t') |V(t')| \phi_{nk}(t')\rangle
  e^{-i\epsilon_{nk}t'+i \epsilon_{n' k'}t'}\nonumber\\
  & = -i\sum_{n' k'}e^{-i\epsilon_{n k}t}|\phi_{n'k'}(t)\rangle\int_0^t dt' \langle \phi_{n' k'}(t') |V(t')| \phi_{nk}(t')\rangle
  e^{-i\epsilon_{nk}(t'-t)+i \epsilon_{n' k'}(t'-t)}.
\end{align}
Above we have used that the Floquet states constitute a complete set of states, and we have employed the identity $U(0,t)U(t,t') = U(0,t')$.

Note that to $O(B)$ the change in the quasi-energy is given by $\delta \epsilon_{n,k}= \langle \phi_{n,k}|V(t)|\phi_{n,k}\rangle$. Since $V(t)$ is proportional to the velocity operator, and the average velocity vanishes in an an eigenstate, 
$\langle \phi_{n,k}|V(t)|\phi_{n,k}\rangle=0$.
Therefore the change in the quasimode comes entirely from the change in the Floquet eigenstate, i.e., $ \delta |\phi_{n k}(t)\rangle  = e^{i\epsilon_{n,k}t}\delta |\psi_{n k}(t)\rangle$ to $O(B)$. 
Substituting for $V(t)$, the change in the quasimode to $O(B)$ is
\begin{align}
   \delta |\phi_{n k}(t)\rangle &=\frac{B}{2q}
  \sum_{n'}  |\phi_{n'k+q}(t)\rangle\int_0^t dt' \langle \phi_{n' k+q}(t') |v^x(k+q/2,t')| \phi_{nk}(t')\rangle
  e^{-i\epsilon_{nk}(t'-t)+i \epsilon_{n' k+q}(t'-t)}\nonumber\\
  &-\frac{B}{2q}
  \sum_{n'}  |\phi_{n'k-q}(t)\rangle\int_0^t dt' \langle \phi_{n' k-q}(t') |v^x(k-q/2,t')| \phi_{nk}(t')\rangle
  e^{-i\epsilon_{nk}(t'-t)+i \epsilon_{n' k-q}(t'-t)}.
\end{align}
Above,
\begin{align}
v^x_{k+q/2}(t)={v}^x(k_x, k_y+q/2,t) = \frac{\partial}{\partial k_x}H(k_x, k_y+q/2,t).
\end{align}
Thus the change in the average quasi-energy to $O(B)$ is
\begin{align}
  \delta E (t) &= \sum_{n,k}f_{nk}\biggl[\langle \phi_{nk}(t)|H_F|\delta \phi_{nk}(t)\rangle + c.c.\biggr]\nonumber\\
  &=\frac{B}{2q}\sum_{n,n',k}f_{nk}\biggl[
   \langle \phi_{nk}(t)|H_F|\phi_{n'k+q}(t)\rangle\int_0^t dt' \langle \phi_{n' k+q}(t') |v^x_{k+q/2}(t')| \phi_{nk}(t')\rangle
  e^{-i\epsilon_{nk}(t'-t)+i \epsilon_{n' k+q}(t'-t)}\nonumber\\
  &- \langle \phi_{nk}(t)|H_F |\phi_{n'k-q}(t)\rangle\int_0^t dt' \langle \phi_{n' k-q}(t') |v^x_{k- q/2}(t')| \phi_{nk}(t')\rangle
  e^{-i\epsilon_{nk}(t'-t)+i \epsilon_{n' k-q}(t'-t)}
  \nonumber \\
  &+ \langle \phi_{n' k+q}(t)|H_F |\phi_{n k}(t)\rangle\int_0^t dt' \langle \phi_{n k}(t') |v^x_{k+q/2}(t')| \phi_{n'k+q}(t')\rangle
  e^{i\epsilon_{nk}(t'-t)-i \epsilon_{n' k+q}(t'-t)}\nonumber\\
  &- \langle \phi_{n'k-q}(t)|H_F |\phi_{n k}(t)\rangle\int_0^t dt' \langle \phi_{n k}(t') |v^x_{k-q/2}(t')| \phi_{n'k-q}(t')\rangle
  e^{i\epsilon_{nk}(t'-t)-i \epsilon_{n' k-q}(t'-t)}\biggr].\label{bsa}
\end{align}
In second and last terms we find it convenient to shift
$k\rightarrow k+q$, and interchange $n,n'$. This gives
\begin{align}
  \delta E (t) =\frac{B}{2q}\sum_{n,n',k}\left(f_{nk}-f_{n',k+q}\right)&\biggl[
   \langle \phi_{nk}(t)|H_F |\phi_{n'k+q}(t)\rangle\int_0^t dt' \langle \phi_{n' k+q}(t') |v^x_{k+q/2}(t')| \phi_{nk}(t')\rangle
   e^{-i\epsilon_{nk}(t'-t)+i \epsilon_{n' k+q}(t'-t)}\nonumber\\
   &+ c.c \biggr] = B M(t).\label{bs2}
\end{align}
Note that the term multiplying $B/(2q)$ is $O(q)$ because it is a product of a  term which is symmetric under exchange of $n,n'$ when $q=0$ (term in square brackets for $q=0$) and another which is anti-symmetric under exchange of $n,n'$
when $q=0$ (the term $\left(f_{n,k}-f_{n',k}\right)$). Thus the prefactor of $B/2q$ vanishes when $q=0$. Now the goal is to expand this prefactor to $O(q)$ and determine $M(t)$. When $n=n'$, the $O(q)$ correction comes from Taylor expanding $(f_{nk}-f_{n k+q})$ to $O(q)$, while for
the $n\neq n'$ term, the $O(q)$ correction comes from Taylor expanding $\langle \phi_{nk}(t)|H_F |\phi_{n'k+q}(t)\rangle$. Thus, in what follows we may replace $v^x_{k+q/2}$ by $v^x_k$.

We introduce the following notation for the Fourier transform of matrix elements between Floquet quasimodes
\begin{align}
  \left[\hat{O}\right]^{m}_{n k,n' k+q} &= \frac{1}{T}\int_0^T dt' e^{-i m \Omega t'} \langle \phi_{n k}(t') |\hat{O}| \phi_{n' k+q}(t')\rangle.
\end{align}
In Eq.~\eqref{bs2} we perform the $t'$ integral by including
a small imaginary part, and extending the integral to $\infty$. This gives the following expression for the average orbital magnetization
\begin{align}
\overline{M}=\lim \limits_{q\rightarrow 0}\frac{1}{2q}\sum_{n,n',k,m}\biggl[
    \left(f_{nk}-f_{n' k+q}\right)\left[H_F\right]^{m}_{n k,n' k+q}\left[v^x(k)\right]^{-m}_{n' k+q,n k}
\frac{i}{\epsilon_{n k}+m \Omega - \epsilon_{n',k+q}} + c.c.\biggr],\label{Mfl1}
\end{align}
where,
\begin{align}
  \left[H_F\right]^{m}_{n k,n' k+q}&= \frac{1}{T}\int_0^T dt' e^{-i m \Omega t'} \langle \phi_{n k }(t') |H_F| \phi_{n' k+q}(t')\rangle\nonumber\\
  &=\frac{1}{2}
  \left[\epsilon_{n',k+q}+\epsilon_{n,k} + m \Omega\right]\sum_{m_1}\langle \phi_{n k}^{(m_1-m)} | \phi_{n' k+q}^{(m_1)}\rangle,
\end{align}
and,
\begin{align}
  \left[v^x(k)\right]^{-m}_{n'k+q,nk} &= \frac{1}{T}\int_0^T dt' e^{i m \Omega t'} \langle \phi_{n' k+q}(t') |\partial_{k_x}H(t')| \phi_{nk}(t')\rangle \nonumber\\
  &= \frac{1}{T}\int_0^T dt' e^{i m \Omega t'} \langle \phi_{n' k+q}(t') |\partial_{k_x}\biggl[H(t')| \phi_{nk}(t')\rangle
    \biggr] - \frac{1}{T}\int_0^T dt' e^{i m \Omega t'} \langle \phi_{n' k+q}(t') |H(t')|\partial_{k_x} \phi_{nk}(t')\rangle\nonumber\\
  & = \frac{1}{T}\int_0^T dt' e^{i m \Omega t'} \langle \phi_{n' k+q}(t') |\partial_{k_x}\biggl[\epsilon_{nk}+i\partial_{t'}| \phi_{nk}(t')\rangle
    \biggr]\nonumber\\
  &  - \frac{1}{T}\int_0^T dt' e^{i m \Omega t'}\biggl[ \epsilon_{n'k+q}\langle \phi_{n' k+q}(t') |\partial_{k_x} \phi_{nk}(t')\rangle
    -i \langle \partial_{t'}\phi_{n' k+q}(t') |\partial_{k_x} \phi_{nk}(t')\rangle\biggr]\nonumber\\
  &= \left(\partial_{k_x}\epsilon_{nk}\right) \frac{1}{T}\int_0^T dt' e^{i m \Omega t'} \langle \phi_{n' k+q}(t')| \phi_{nk}(t')\rangle \nonumber\\
&  + \frac{1}{T}\int_0^T dt' e^{i m \Omega t'}\left(\epsilon_{nk}-\epsilon_{n'k+q}+i\partial_{t'}\right)
  \langle \phi_{n' k+q}(t')|\partial_{k_x} \phi_{nk}(t')\rangle\nonumber\\
  &=\left(\partial_{k_x}\epsilon_{nk}\right)\sum_{m_1}\langle \phi_{n' k+q}^{(m+m_1)} |\phi_{n k}^{(m_1)}\rangle
  + \biggl(\epsilon_{n k}-\epsilon_{n' k+q}+m \Omega\biggr)\sum_{m_1}\langle \phi_{n' k+q}^{(m+m_1)} |\partial_{k_x}\phi_{nk}^{(m_1)}\rangle.
\end{align}
Eq.~\eqref{Mfl1} corresponds to Eq.~\eqref{Mfl}.

In order to Taylor expand in $q$, it is convenient to separate the terms into $n\neq n'$ and $n=n'$ as follows
\begin{align}
\overline{M}&= \lim\limits_{q\rightarrow 0}\frac{1}{2q}\sum_{n\neq n',k,m}\biggl[
    \left(f_{nk}-f_{n' k+q}\right)\left[H_F\right]^{m}_{n k,n' k+q}\left[v^x(k)\right]^{-m}_{n' k+q,n k}
    \frac{i}{\epsilon_{n k}+m \Omega - \epsilon_{n',k+q}} + c.c.\biggr]\nonumber\\
&+ \lim\limits_{q\rightarrow 0}\frac{1}{2q}\sum_{n,k,m}\biggl[
    \left(f_{nk}-f_{n k+q}\right)\left[H_F\right]^{m}_{n k,n k+q}\left[v^x(k)\right]^{-m}_{n k+q,n k}
    \frac{i}{\epsilon_{n k}+m \Omega - \epsilon_{n,k+q}} + c.c.\biggr].
\end{align}
Substituting for matrix elements of $v_x$ and $H_F$ we obtain
\begin{align}
\overline{M}&= \lim\limits_{q\rightarrow 0}\frac{1}{2q}\sum_{n\neq n',k,m}\left(f_{nk}-f_{n' k+q}\right)\biggl[
   \biggl\{\frac{1}{2}
   \left[\epsilon_{n',k+q}+\epsilon_{n,k}+m \Omega\right]\sum_{m_1}\langle \phi_{n k}^{(m_1-m)} | \phi_{n' k+q}^{(m_1)}\rangle\biggr\}\nonumber\\
  &\times\biggl\{\left(\partial_{k_x}\epsilon_{nk}\right)\sum_{m_2}\langle \phi_{n' k+q}^{(m+m_2)} |\phi_{n k}^{(m_2)}\rangle
  + \biggl(\epsilon_{n k}-\epsilon_{n' k+q}+m \Omega\biggr)\sum_{m_2}\langle \phi_{n' k+q}^{(m+m_2)} |\partial_{k_x}\phi_{nk}^{(m_2)}\rangle\biggr\}
    \frac{i}{\epsilon_{n k}+m \Omega - \epsilon_{n',k+q}} + c.c.\biggr]\nonumber\\
&+ \lim\limits_{q\rightarrow 0}\frac{1}{2q}\sum_{n,k,m}\left(f_{nk}-f_{n k+q}\right)\biggl[
  \biggl\{\frac{1}{2}
   \left[\epsilon_{n,k+q}+\epsilon_{n,k}+ m\Omega \right]\sum_{m_1}\langle \phi_{n k}^{(m_1-m)} | \phi_{n,k+q}^{(m_1)}\rangle\biggr\}\nonumber\\
  &\times\biggl\{\left(\partial_{k_x}\epsilon_{nk}\right)\sum_{m_2}\langle \phi_{n k+q}^{(m+m_2)} |\phi_{n k}^{(m_2)}\rangle
  + \biggl(\epsilon_{n k}-\epsilon_{n k+q}+m \Omega\biggr)\sum_{m_2}\langle \phi_{n k+q}^{(m+m_2)} |\partial_{k_x}\phi_{nk}^{(m_2)}\rangle\biggr\}
    \frac{i}{\epsilon_{n k}+m \Omega - \epsilon_{n,k+q}} + c.c.\biggr].
\end{align}

When $n\neq n'$, the term in the first line above $\sum_{m_1}\langle \phi_{n k}^{(m_1-m)} | \phi_{n' k+q}^{(m_1)}\rangle$ is such that when $q=0$, this term vanishes, i.e,  $\sum_{m_1}\langle \phi_{n k}^{(m_1-m)} | \phi_{n' k}^{(m_1)}\rangle=0$. This is because $\langle \phi_n(t)|\phi_{n'}(t)\rangle=0$ and therefore all the Fourier components of this overlap should be zero.
Thus for the $n\neq n'$ term, we need to Taylor expand only $\sum_{m_1}\langle \phi_{n k}^{(m_1-m)} | \phi_{n' k+q}^{(m_1)}\rangle$ to $O(q)$.
In addition, for the $n=n'$ term, the expression
\begin{align}
&\sum_{m_1}\langle \phi_{n k}^{(m_1-m)} | \phi_{n k+q}^{(m_1)}\rangle \times \sum_{m_2}\langle \phi_{n k+q}^{(m+m_2)} |\phi_{n k}^{(m_2)}\rangle \nonumber\\
&= \biggl[\frac{1}{T}\int_0^T dt e^{-i m\Omega t}\langle \phi_{n k}(t) | \phi_{n k+q}(t)\rangle\biggr]\biggl[\frac{1}{T}\int_0^T dt' e^{i m \Omega t'} \langle \phi_{n k+q}(t') |\phi_{n k}(t')\rangle\biggr],
\end{align}
and is purely real. Thus the term proportional to $\left(\partial_{k_x}\epsilon_{nk}\right)$ does not contribute due to the $i$ factor multiplying it.
This leads to 
\begin{align}
\overline{M}&= \frac{1}{2}\sum_{n\neq n',k,m}\left(f_{nk}-f_{n' k}\right)\biggl[
   \biggl\{\frac{1}{2}
   \left[\epsilon_{n',k}+\epsilon_{n,k}+ m\Omega \right]\sum_{m_1}\langle \phi_{n k}^{(m_1-m)} | \partial_{k_y}\phi_{n' k}^{(m_1)}\rangle\biggr\}
   \biggl\{\sum_{m_2}\langle \phi_{n' k}^{(m+m_2)} |\partial_{k_x}\phi_{nk}^{(m_2)}\rangle\biggr\}i + c.c.\biggr]\nonumber\\
&+ \lim\limits_{q\rightarrow 0}\frac{1}{2q}\sum_{n,k,m}\left(f_{nk}-f_{n k+q}\right)\biggl[
  \biggl\{\frac{1}{2}
  \left[\epsilon_{n,k+q}+\epsilon_{n,k}+ m\Omega \right]\sum_{m_1}\langle \phi_{n k}^{(m_1-m)} | \phi_{n k+q}^{(m_1)}\rangle\biggr\}
  \biggl\{i\sum_{m_2}\langle \phi_{n k+q}^{(m+m_2)} |\partial_{k_x}\phi_{nk}^{(m_2)}\rangle\biggr\} + c.c.\biggr].\label{Mint3}
\end{align}
As before, since $\langle \phi(t)| \phi(t)\rangle=1$ the term $\sum_{m_1}\langle \phi_{n k}^{(m_1-m)} | \phi_{n k+q}^{(m_1)}\rangle = \delta_{m=0} + O(q)\delta_{m\neq 0}$. This is because, 
when $q=0$, all time-dependence should vanish (i.e, $m=0$). So to leading order in $q$, only the $m=0$ term in the second line survives, giving
\begin{align}
\overline{M}&= \frac{1}{2}\sum_{n\neq n',k,m}\left(f_{nk}-f_{n' k}\right)\biggl[
   \biggl\{\frac{1}{2}
   \left[\epsilon_{n',k}+\epsilon_{n,k}+m\Omega\right]\sum_{m_1}\langle \phi_{n k}^{(m_1-m)} | \partial_{k_y}\phi_{n' k}^{(m_1)}\rangle\biggr\}
   \biggl\{\sum_{m_2}\langle \phi_{n' k}^{(m+m_2)} |\partial_{k_x}\phi_{nk}^{(m_2)}\rangle\biggr\}i + c.c.\biggr]\nonumber\\
&+ \lim\limits_{q\rightarrow 0}\frac{1}{2q}\sum_{n,k}\left(f_{nk}-f_{n k+q}\right)\frac{1}{2}
  \left[\epsilon_{n,k+q}+\epsilon_{n,k}\right]
   \biggl[i\sum_{m_1}\langle \phi_{n k+q}^{(m_1)} |\partial_{k_x}\phi_{nk}^{(m_1)}\rangle + c.c.\biggr].\label{Mint4}
\end{align}

Now we divide and multiply by $\left(\epsilon_{nk}-\epsilon_{n,k+q}\right)$ in the second term and use the identity
\begin{align}
&  (\epsilon_{nk}-\epsilon_{n,k+q})  \sum_{m_1}\langle \phi_{n k+q}^{(m_1)} |\partial_{k_x}\phi_{nk}^{(m_1)}\rangle
  = \overline{\langle \phi_{n k+q}| \epsilon_{nk}-H_F|\partial_{k_x}\phi_{nk}\rangle}
  \approx q \overline{\langle \partial_{k_y}\phi_{n k}| \epsilon_{nk}-H_F|\partial_{k_x}\phi_{nk}\rangle},
\end{align}
to obtain
\begin{align}
\overline{M}&= \frac{1}{2}\sum_{n\neq n',k,m}\left(f_{nk}-f_{n' k}\right)\biggl[
   \biggl\{\frac{1}{2}
   \left[\epsilon_{n',k}+\epsilon_{n,k}+ m\Omega\right]\sum_{m_1}\langle \phi_{n k}^{(m_1-m)} | \partial_{k_y}\phi_{n' k}^{(m_1)}\rangle\biggr\}
   \biggl\{\sum_{m_2}\langle \phi_{n' k}^{(m+m_2)} |\partial_{k_x}\phi_{nk}^{(m_2)}\rangle\biggr\}i + c.c.\biggr]\nonumber\\
&+ \frac{1}{2}\sum_{n,k}\biggl\{\lim\limits_{q\rightarrow 0}\frac{\left(f_{nk}-f_{n k+q}\right)}{\epsilon_{nk}-\epsilon_{n,k+q}}\biggr\}
\epsilon_{n,k}\biggl[i\overline{\langle \partial_{k_y}\phi_{n k}| \epsilon_{nk}-H_F|\partial_{k_x}\phi_{nk}\rangle}+c.c\biggr].
\label{Mint5}
\end{align}

Since $\langle \phi_{n',k}(t)|\phi_{n,k}(t)\rangle=$ constant in time,
$\partial_{k_i}\left[\langle \phi_{n',k}(t)|\phi_{n,k}(t)\rangle\right]=0$.
This must hold for each Fourier component giving $\partial_{k_i}\left[\sum_{m_1}\langle \phi_{n',k}^{(m_1-m)}|\phi_{n,k}^{(m_1)}\rangle\right]=0$ for each $m$. Using this, we can move the derivatives between the bras and kets to show that
\begin{align}
  \sum_{n\neq n',k,m}\left(f_{nk}-f_{n' k}\right)m\Omega\biggl[\biggl\{\sum_{m_1}\langle \phi_{n k}^{(m_1-m)} | \partial_{k_y}\phi_{n' k}^{(m_1)}\rangle\biggr\}
   \biggl\{\sum_{m_2}\langle \phi_{n' k}^{(m+m_2)} |\partial_{k_x}\phi_{nk}^{(m_2)}\rangle\biggr\}i + c.c.\biggr]   = 0.
\end{align}
Similar manipulations lead to
\begin{align}
  \overline{M} &=-\frac{1}{2}{\rm Im}\biggl[\sum_{n,n',k,m}f_{nk}\left(\epsilon_{n',k}+\epsilon_{n,k}\right)
   \biggl\{\sum_{m_1}\langle \partial_{k_y} \phi_{n k}^{(m_1-m)} | \phi_{n' k}^{(m_1)}\rangle\biggr\}
   \biggl\{\sum_{m_2}\langle \phi_{n' k}^{(m+m_2)} |\partial_{k_x}\phi_{nk}^{(m_2)}\rangle\biggr\}\biggr]\nonumber\\
   & + \frac{1}{2}{\rm Im}\biggl[\sum_{n,n',k,m}f_{n'k}\left(\epsilon_{n',k}+\epsilon_{n,k}\right)
   \biggl\{\sum_{m_1}\langle \phi_{n k}^{(m_1-m)} | \partial_{k_y}\phi_{n' k}^{(m_1)}\rangle\biggr\}
   \biggl\{\sum_{m_2}\langle \partial_{k_x}\phi_{n' k}^{(m+m_2)} |\phi_{nk}^{(m_2)}\rangle\biggr\}\biggr]\\
  &+{\rm Im} \biggl[\sum_{n,k}\biggl\{\lim\limits_{q\rightarrow 0}\frac{\left(f_{nk}-f_{n k+q}\right)}{\epsilon_{nk}-\epsilon_{n,k+q}}\biggr\}\epsilon_{n,k}
  \overline{\langle \partial_{k_y}\phi_{n k}(t) | \epsilon_{nk}- H_F |\partial_{k_x}\phi_{nk}(t)\rangle}\biggr].
\end{align}
The above can be written as
\begin{align}
  \overline{M} &=-\frac{1}{2}{\rm Im}\biggl[\sum_{n,n',k}f_{nk}
   \overline{\langle \partial_{k_y} \phi_{n k}(t) |\epsilon_{nk}+H_F| \phi_{n' k}(t)\rangle
   \langle \phi_{n' k}(t) |\partial_{k_x}\phi_{nk}(t)\rangle}\biggr]\nonumber\\
   & + \frac{1}{2}{\rm Im}\biggl[\sum_{n,n',k}f_{n'k}
    \overline{\langle \partial_{k_x}\phi_{n' k}(t)|\epsilon_{n'k}+H_F|\phi_{nk}(t)\rangle
      \langle \phi_{n k}(t) | \partial_{k_y}\phi_{n' k}(t)\rangle}\biggr]\\
  &+{\rm Im} \biggl[\sum_{n,k}\biggl\{\lim\limits_{q\rightarrow 0}\frac{\left(f_{nk}-f_{n k+q}\right)}{\epsilon_{nk}-\epsilon_{n,k+q}}\biggr\}\epsilon_{n,k}
  \overline{\langle \partial_{k_y}\phi_{n k}(t) | \epsilon_{nk}- H_F |\partial_{k_x}\phi_{nk}(t)\rangle}\biggr].
\end{align}
Now we remove the complete set of states,
$\sum_n |\phi_{nk}(t)\rangle
\langle \phi_{n k}(t) |=1$, and by noting that the second line is simply the complex conjugate of the first, we obtain
\begin{align}
  \overline{M} &=-{\rm Im}\biggl[\sum_{n,k}f_{nk}
   \overline{\langle \partial_{k_y} \phi_{n k}(t) |\epsilon_{nk}+H_F|\partial_{k_x}\phi_{nk}(t)\rangle}\biggr]\nonumber\\
  &+{\rm Im} \biggl[\sum_{n,k}\biggl\{\lim\limits_{q\rightarrow 0}\frac{\left(f_{nk}-f_{n k+q}\right)}{\epsilon_{nk}-\epsilon_{n,k+q}}\biggr\}\epsilon_{n,k}
  \overline{\langle \partial_{k_y}\phi_{n k}(t) | \epsilon_{nk}- H_F |\partial_{k_x}\phi_{nk}(t)\rangle}\biggr].
\end{align}

Multiplying by $-e/\hbar$, and writing ${\rm Im}(a) = (a-a^*)/2i$, we obtain
\begin{align}
  \overline{M} &= -\frac{e}{2\hbar}{\rm Im}\sum_{n,k} \biggl[f_{nk}
    \overline{\langle \partial_{\vec{k}} \phi_{n k}(t) |\left(\epsilon_{nk}+H_F\right)
      \times |\partial_{\vec{k}}\phi_{nk}(t)\rangle}\nonumber\\
  &-\biggl\{\lim\limits_{q\rightarrow 0}\frac{\left(f_{nk}-f_{n k+q}\right)}{\epsilon_{nk}-\epsilon_{n,k+q}}\biggr\}\epsilon_{n,k}
  \overline{\langle \partial_{\vec{k}}\phi_{n k}(t) | \left(\epsilon_{nk}- H_F\right)  \times |\partial_{\vec{k}}\phi_{nk}(t)\rangle}\biggr].
\end{align}
The above is Eq.~\eqref{M}.

\subsection{Simplified formulas for the two-band model}

When there are only two bands $n=u,d$ as in the case of periodically driven graphene, the formulas for the orbital magnetization simplify. For
the orbital magnetization under assumptions of thermal equilibrium, and taking into account corrections from $E-\mu \langle N\rangle$, we have
\begin{align}\label{M2band}
  \overline{M}
  & =-\frac{e}{2\hbar}{\rm Im}\sum_{k}\biggl[\left(f_{dk}-f_{uk}\right)\left(\epsilon_{d,k}+\epsilon_{u,k}-2\mu\right)\overline{F_{xy}(k,t)}\nonumber\\
&-(\epsilon_{d,k}-\epsilon_{u,k})\overline{F_{xy}(k,t)}\sum_{n=d,u}f'_{nk}(\epsilon_{nk}-\mu)\biggr].
\end{align}

For the orbital magnetization for a quench occupation of the bands, and looking at corrections only to $E$, we have
\begin{align}
 \overline{M}_Q
  & =-\frac{e}{2\hbar}{\rm Im}\sum_{k}\biggl[\left(f_{dk}-f_{u k}\right)\left(\epsilon_{d,k}+\epsilon_{u,k}\right)\overline{F_{xy}(k,t)}\\
&-\left(\epsilon_{d,k}-\epsilon_{u,k}\right)\overline{F_{xy}(k,t)}\sum_{\alpha}\biggl\{  |\langle \phi_{dk}|\psi^{\rm in}_{\alpha k}\rangle|^2 \partial_{\epsilon}f^{\rm in}_{\alpha} \frac{v_y^{\rm in}(\alpha, k)}{v_y(d,k)}\epsilon_{d,k}+ \langle \phi_{uk}|\psi^{\rm in}_{\alpha k}\rangle|^2 \partial_{\epsilon}f^{\rm in}_{\alpha} \frac{v_y^{\rm in}(\alpha, k)}{v_y(u,k)} \epsilon_{u,k}\nonumber\\
&+f^{\rm in}_{\alpha,k}   \frac{\partial_{k_y}\left(|\langle \phi_{dk}|\psi^{\rm in}_{\alpha k}\rangle|^2\right)}{v_y(d,k)} \epsilon_{d,k} + f^{\rm in}_{\alpha,k}   \frac{\partial_{k_y}\left(|\langle \phi_{uk}|\psi^{\rm in}_{\alpha k}\rangle|^2\right)}{v_y(u,k)} \epsilon_{u,k}\biggr\}\biggr].
\end{align}
Above $\overline{F_{xy}}$ is the time-average of the Berry curvature and $\alpha$ are the bands of the system before the quench. 
Note that when there is particle-hole symmetry $\epsilon_{d,k} =-\epsilon_{u,k}$, then the first term in the above equations proportional to
$\epsilon_{d,k}+\epsilon_{u,k}$ does not contribute. In addition, when the system is in thermal equilibrium at zero temperature, with one band fully occupied, and the other empty, then Eq.~\eqref{M2band} gives 
$\overline{M}/A = -(e/\hbar)C \mu/2\pi$.

\section{Van-Hove Singularities}
For the sake of completeness, in Fig.~\ref{fig:S1} we show the same data as in Fig.~\ref{fig:2}(d) but with the full magnetization range.
\begin{figure*}
    \centering
    \includegraphics[width=\textwidth]{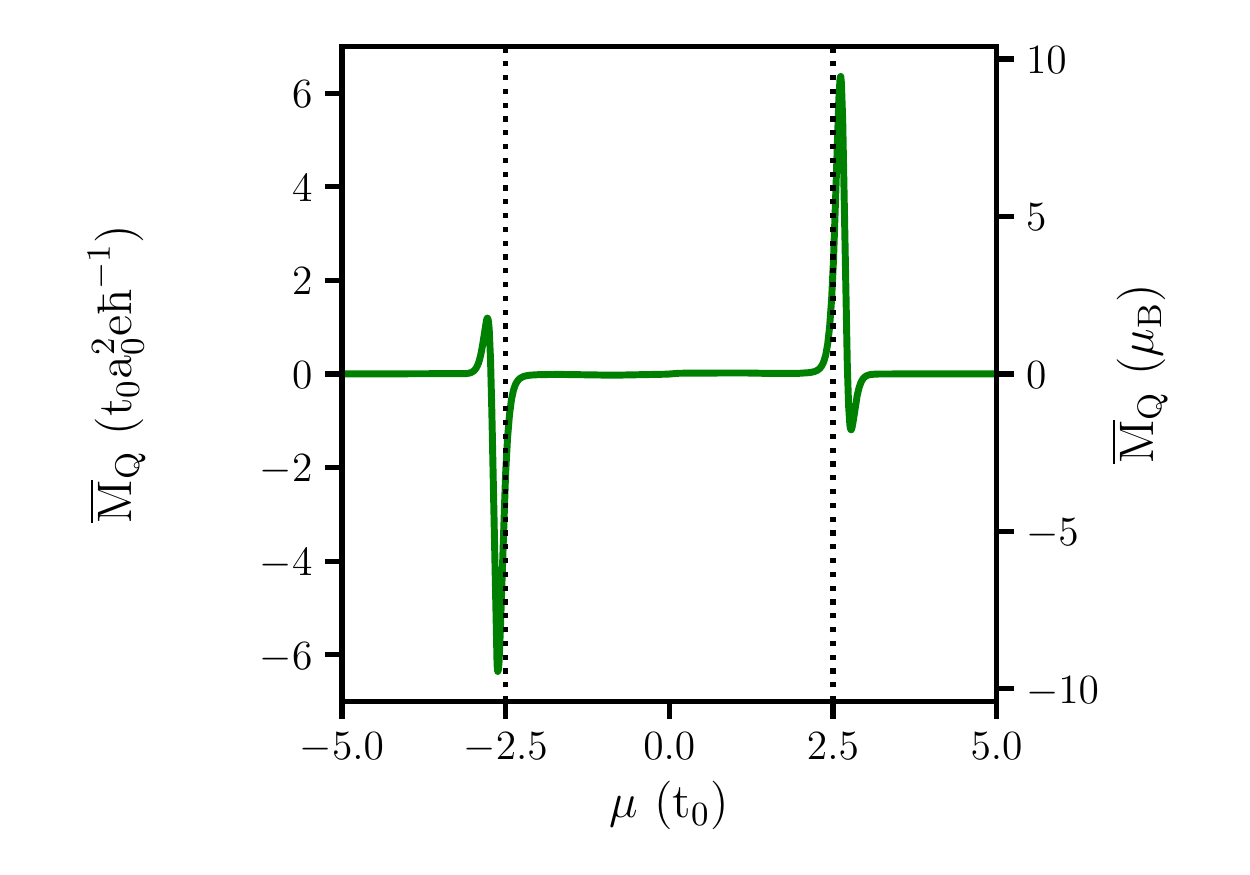}
    \caption{Same as Fig.~\ref{fig:2}(d) i.e, the orbital magnetization for a quench occupation of the bands for $C=3$, but with the full magnetization range shown.}
    \label{fig:S1}
\end{figure*}
Fig.~\ref{fig:S2} shows the Floquet band structure within the 1st Floquet Brillouin zone along the Dirac point for the three driving cases with $C=1$ (a), $C=3$ (c), and $C=0$ (e). For better orientation the static bands are shown by black-dashed lines. The inset in (a) shows the Dirac point with the light-induced gap opening. The corresponding density of states is presented in subpanels (b), (d), and (f).      
\begin{figure*}
    \centering
    \includegraphics[width=\textwidth]{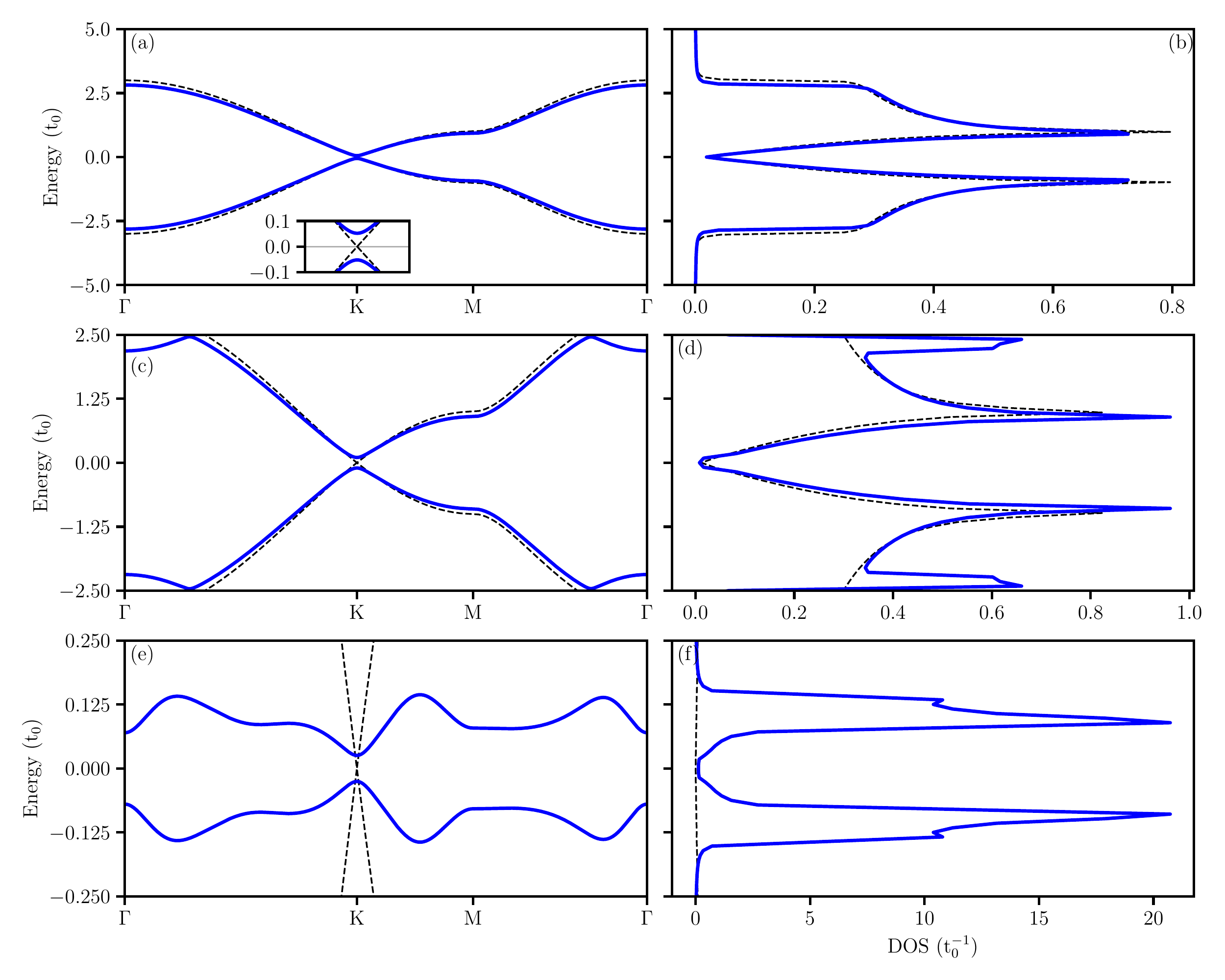}
    \caption{Floquet bands and Brillouin-zone averaged density of states (DOS) for the three different driving cases with Chern number $C=1$ (a,b), $C=3$ (c,d), and $C=0$ (e,f) within the first Floquet Brillouin zone, for periodic boundary conditions in $x$ and $y$. Black dashed lines indicate the respective values for the static case (undriven graphene). The inset in (a) shows the driving induced gap. We assume a Lorentzian broadening of the energy levels with a width of $0.05t_0^{-1}$ (b), $0.025t_0^{-1}$ (d), and $0.005t_0^{-1}$ (f). Numerical parameters such as number of Floquet harmonics and $k$-grid size are the same as for Fig.~\ref{fig:2}.}
    \label{fig:S2}
\end{figure*}

\section{Orbital magnetization density for the quench}
In Fig.~\ref{fig:S3} we present the $k$-resolved orbital magnetization density after a sudden quench, as described by Eq.~\eqref{MQ}, for the driving schemes corresponding to  $C=3$ (a) and $C=0$ (b). We use the same parameters for $\mu$ and $\beta^{-1}$ as for Fig.~\ref{fig:3}, where $\mu$ and $\beta^{-1}$ now characterize the state before the quench, i.e., the temperature and occupation of undriven graphene. Due to the non-equilibrium occupations, the magnetization shows more complexity, for example, clouds of non-zero magnetization density at the Dirac point position (a). In contrast, for thermal equilibrium occupation of the bands, and for the same chemical potential, the magnetization density at the Dirac points is zero (c.f. Fig.~\ref{fig:3}a).    
\begin{figure*}
    \centering
    \includegraphics[width=\textwidth]{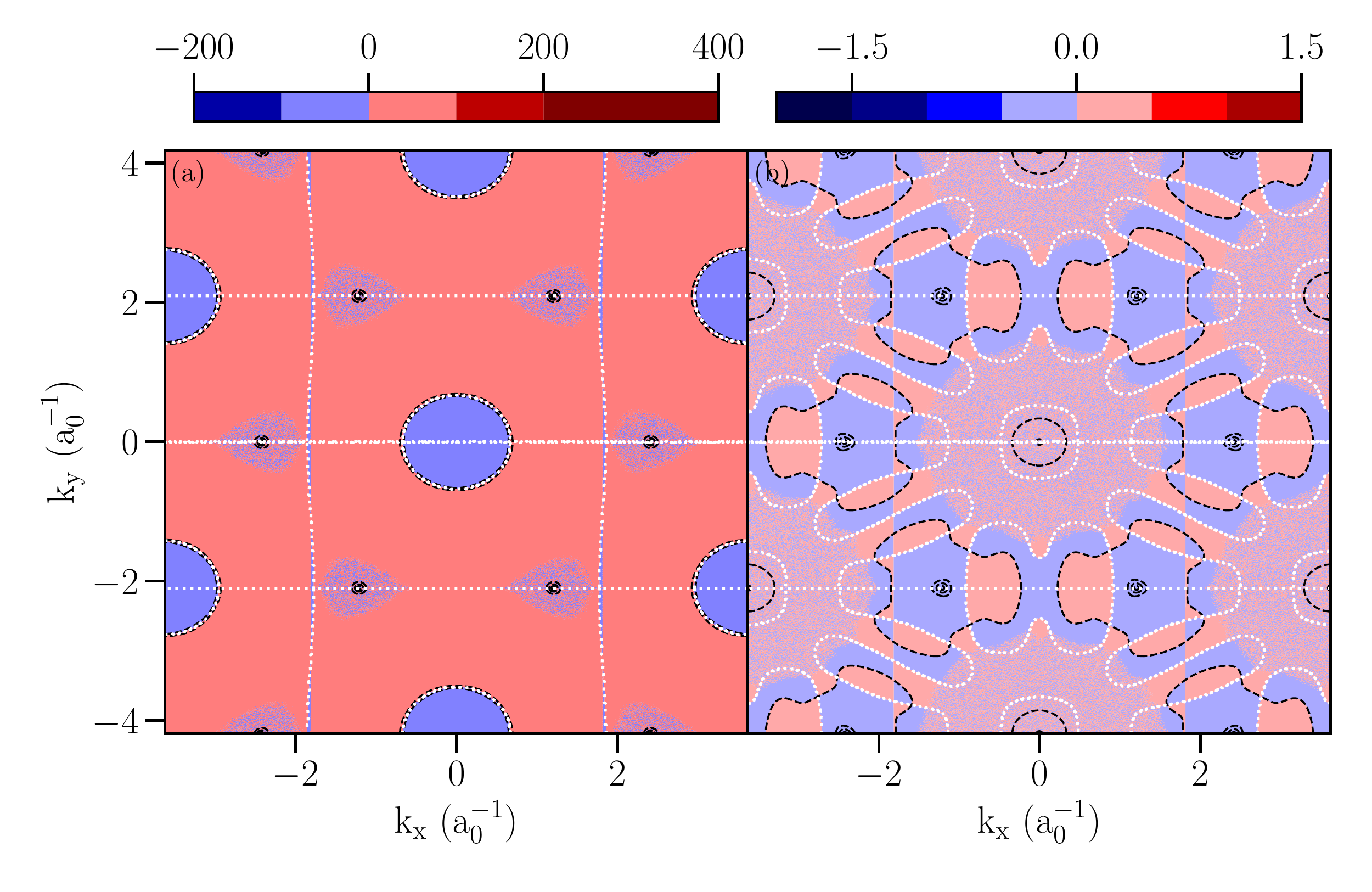}
    \caption{The orbital magnetization density, but with occupations according to a sudden quench (Eq.~\eqref{fq}) and with the magnetization density corresponding to the integrand  of Eq.~\eqref{MQ}. Black contours in both panels indicate regions where the Berry curvature has significant contributions. White dots in both panels indicate regions with vanishing Floquet band velocity in the $y$-direction, and therefore are related to the vHs. The initial state i.e, the state of undriven graphene is taken to be in thermal equilibrium at a temperature $\beta^{-1}=0.05t_0$ and a chemical potential $\mu =2.3 t_0$ (a) and $\mu=0.1 t_0$ (b). Grid size and number of Floquet harmonics are the same as Fig.~\ref{fig:3}.}
    \label{fig:S3}
\end{figure*}

\end{widetext}

\end{document}